\documentclass[aps,showpacs,prstper,citeautoscript,reprint]{revtex4-1}

\usepackage[bf, tiny, center, compact, uppercase]{titlesec}
\usepackage{array}
\usepackage{amsmath}
\usepackage{units}
\usepackage{graphicx}
\usepackage{braket}
\usepackage{xfrac}
\usepackage{enumitem}
\usepackage[normalem]{ulem}
\usepackage{booktabs}
\usepackage{tabularx}
\usepackage{multirow, bigstrut}
\usepackage{hhline}

\titlespacing\section{0pt}{12pt plus 1pt minus 2pt}{6pt plus 1pt minus 2pt}

\usepackage{times}
\newcommand{\op}[1]{\ensuremath{\hat{#1}}}

\setcitestyle{numbers}
\setcitestyle{square}

\begin{document}
\setlength{\abovedisplayskip}{1pt}
\setlength{\belowdisplayskip}{1pt}
\setlength{\abovedisplayshortskip}{0pt}
\setlength{\belowdisplayshortskip}{0pt}


\title{The effect of giving explicit incentives to correct mistakes on subsequent problem solving in quantum mechanics}

\author{Benjamin R. Brown}
\author{Chandralekha Singh}
\affiliation{Department of Physics and Astronomy, University of Pittsburgh, Pittsburgh, PA 15260}

\author{Andrew Mason}
\affiliation{University of Central Arkansas, Conway, AR  72035}
 

\begin{abstract}
One attribute of experts is that they are likely to learn from their own mistakes.  Experts are unlikely to make the same mistakes when asked to solve a problem a second time, especially if they have had access to a correct solution. Here, we discuss a study spanning several years in which advanced undergraduate physics students in a quantum mechanics course were given identical problems in both the midterm exam and final exam.  Approximately half of the students were given incentives to correct their mistakes in the midterm exam and they could get back up to 50\% of the points lost on each midterm exam problem. The solutions to the midterm exam problems were provided to all students in both groups but those who corrected their mistakes were provided the solution after they submitted their corrections to the instructor.  The performance on the final exam on the same problems suggests that students who were given incentives to correct their mistakes significantly outperformed those who were not given an incentive. The incentive to correct the mistakes  had the greatest impact on the final exam performance of students who performed poorly on the midterm exam. 
\end{abstract}

\maketitle
\section{Introduction}
One characteristic of experts is that they are likely to use problem solving as an opportunity for learning \cite{reif,chi,ericsson}. In particular, experts are likely to reflect upon their mistakes in their problem solution automatically in order to repair, extend and organize their knowledge structure.  Unfortunately, for many students in physics courses, problem solving is a missed learning 
opportunity \cite{yerushalmi2012,yerushalmi2012ii,mason2010,henderson}. 
Without guidance, many students do not reflect upon the problem solving process after solving problems in order to learn from them, nor do they make an effort to learn from their mistakes after the graded problems are returned to them \cite{yerushalmi2012,yerushalmi2012ii,mason2010,henderson}. 

However, closing the ``performance gap'' between high and low achieving students and ensuring that all students excel in physics courses are important goals of physics education research \cite{mazur1,mazur2}.  Prior research suggests that only providing students worked examples is insufficient \cite{atkinson} and effective approaches to learning involve engaging students in meta-cognition or self-monitoring while they solve problems \cite{meta,meta2}. For example, research suggests that students who went through a productive failure cycle, in which they worked in groups to solve complex ill-structured math problems without any scaffolding support struggled to learn up until a consolidation lecture by the instructor \cite{kapur}. However, these students from the productive failure condition significantly outperformed their counterparts from the lecture and practice condition  on both well- and ill-structured problems on the posttests \cite{kapur}. After the posttest, they also demonstrated significantly better performance in using structured-response scaffolds to solve problems on a new topic not even covered during instruction. Similarly, Schwartz et al. have proposed invention tasks to prepare students for future learning \cite{schwartz}.

It is often implicitly assumed that, unlike students in introductory physics courses, most students in advanced  courses have become independent learners. They will invest time to learn from their own mistakes, even if the instructors do not reward them for fixing their mistakes, e.g., by explicitly asking them to correct their mistakes by giving them grade incentives \cite{yerushalmi2012,yerushalmi2012ii}. Contrary to these beliefs, our earlier investigation found that advanced students in a quantum mechanics course did not automatically improve their performance from midterm to final exam on identical questions even when they were provided the correct solutions and their own graded exams \cite{mason2010}. There was a lack of reflection by supposedly mature students; many students did not make use of mistakes identified on a midterm exam as an opportunity to repair and better organize their knowledge. In individual interviews with some of the students, we also found evidence that even in these advanced courses, many students do not automatically learn from their mistakes and they often resort to rote learning strategies for getting through the course.

These issues are particularly important considering that the diversity in the prior preparation of  students at all levels has increased. Many students need explicit guidance and support not only in the introductory courses, but also in the advanced courses. One instructional strategy that may help is to explicitly prompt students to learn from their mistakes by rewarding them for correcting their mistakes. This activity can also help them learn to make use of problem solving as a learning opportunity.
\section{Goal and Methodology}
Here we report on an investigation focused on how well students in a junior-senior level undergraduate quantum mechanics course learn from their mistakes on midterm exam problems when provided with time and grade incentives for re-submitting written corrected solutions after the graded midterm exams were returned to them. Using a comparison group that did not self-diagnose mistakes on the midterm exams, we investigated the effects of this self-diagnosis of mistakes on subsequent problem solving on some of the same problems repeated on the final exam. 

The study took place over four years with the same instructor and textbook and the course was primarily taught in a traditional lecture format. Students were assigned weekly homework throughout the fifteen-week semester. For each course, there were two midterm exams and a final exam (all exams and homework were identical across the four years). Both midterm exams covered only limited topics while the final exam was comprehensive. Students were instructed in all of the relevant concepts before the exams, and homework was assigned each week from the material covered in that week. The first midterm exam took place approximately eight weeks after the semester started; the second midterm exam took place four weeks after the first midterm exam. For this study, two problems were selected from each of the midterm exams and were repeated verbatim on the final exam along with other problems not asked earlier.

In the first and third year of the study (total 33 students), students were not provided any explicit incentives to learn from their mistakes on the midterm exam (comparison group).  However, in the second and fourth years (total 30 students), students were given a grade incentive to correct their mistakes (incentivized group): they could earn up to 50\% of the points they had lost for the corrections. Students in the incentivized group were directed to work on their own while correcting mistakes, but were free to use any resources including homework, notes, and books for help. Solutions to all of the midterm exam questions were available to students on a course website except that the incentivized group was provided the solutions after the students had submitted their corrected midterm exam solutions. 

Our goal was to explore the extent to which students in each group use their mistakes as a learning opportunity and whether their performance on the same problems administered a second time on the final exam (which we call the posttest) is significantly better than the corresponding performance on the midterm exams (which we call the pretest). In particular, we looked for correlation between the midterm exam score and final exam score on the four common problems for each group (incentivized and comparison groups). 

All problems were graded according to a rubric developed by the researchers \cite{mason2010} with better than 95\% inter-rater reliability for 25\% of the students graded separately (see the summary of the rubric used for problem 1 in Table I).
As shown in Table I, on the problem, each student received an overall score, which is the average of the “Invoking” and “Applying” scores.  Each of these general criteria (invoking and applying) scores is in turn the average of the specific criteria  scores (see Table I), which can take on the values 0, 1, $1/3$, $1/2$, $2/3$, or N/A.  ``N/A" arises in situations when certain criteria are not relevant in determining a score. 

\newlength{\rubricGeneral}
\setlength{\rubricGeneral}{0.2\linewidth}

\begin{table}[th!]
  \setlength{\tabcolsep}{.01\linewidth}

  \abovedisplayskip=6pt
  \belowdisplayskip=6pt
  \abovedisplayshortskip=0pt
  \belowdisplayshortskip=3pt

  \caption[Summary of the Rubric Used for Problem 1.]{Summary of the rubric used for problem 1.  In this problem, students are asked to find the expectation value of an observable $Q$ in terms of eigenstates and eigenvalues of the corresponding operator.   }
  \label{tabIncentiveRubric1}

  \centering
  \begin{tabularx}{\linewidth}{|m{\rubricGeneral}|X|}
  \hline
   {General Criteria} & {Specific Criteria}\\
  \hline


  \multirow{5}{\rubricGeneral}{Invoking appropriate concepts} & 
  {Spectral decomposition expressing identity operator in terms of a complete set of eigenstates $\ket{\psi_n}$:} \\
  & $\op{I} = \sum \ket{\psi_n}\bra{\psi_n}$ \\
  & Or expressing general state in terms of the eigenstates of $\op{Q}$: \\
  & $\ket{\psi} = \sum c_n \ket{\psi_n} \text{, where }c_n = \braket{\psi_n | \psi}$\\
 \cline{2-2}
  & Make use of $\op{Q}\ket{\psi_n} = \lambda_n\ket{\psi_n}$ \bigstrut\\
 \cline{2-2}
  & $\braket{\psi_n | \psi}^{*} = \braket{\psi | \psi_n}$ \\
 \cline{2-2}
  & Using legitimate principles or concepts that are not appropriate in this problem.\\
 \cline{2-2}
  & Using invalid principles or concepts (for instance, confusing a general state ￼ $\ket{\psi}$ with
an eigenstate $\ket{\psi_n}$).\\
\hline
 \multirow{3}{\rubricGeneral}{Applying \newline concepts} & 

Inserting spectral decomposition into the expression for expectation value\\ \cline{2-2}

 & Eigenvalue evaluated and treated as number.\\ \cline{2-2}

 & Probability expressed in terms of $\braket{\psi_n | \psi}$ and $\braket{\psi | \psi_n}$\\
\hline
\end{tabularx}
\end{table}

\section{Results}
\begin{table}[htbp!]
  \caption{Pretest, posttest, and gain for students in the comparison and incentivized groups broken down into low, medium and high performance categories based on students’ pretest score and for all students. While the pretest scores are comparable for the comparison and incentivized groups, the posttest scores are significantly higher for the incentivized group.  The gap between low and high categories shrinks from 42.2\% in the comparison group to 19.9\% in the incentivized group.}
  \label{tabIncentiveGains}
  \centering
  \begin{tabularx}{\linewidth}{>{\itshape}l XXr c XXr}
  \toprule
  
  & \multicolumn{3}{c}{Comparison Group} & &\multicolumn{3}{c}{Incentivized Group} \\
    & \multicolumn{3}{c}{($N=33$)} & &\multicolumn{3}{c}{($N=30$)} \\
  \hhline{~---~---}

  & Pre & Post & Gain& & Pre & Post & Gain \\
  Low &34.6 &50.8 &+16.2&   &33.2 &78.2 &+45.0 \\
  Medium  &64.7 &66.3 &+1.6&    &67.3 &88.5 &+21.3 \\
  High  &96.0 &93.0 &-2.9&    &96.9 &98.1 &+1.2 \\
  \hhline{----~---} 
  
  \multicolumn{1}{l}{\bfseries All}   &{\bfseries 67.9} &{\bfseries 71.5} &{\bfseries +3.6}&    &{\bfseries 67.6} &{\bfseries 88.4} &{\bfseries +20.8}\\
  
  \bottomrule
  \end{tabularx}
\end{table}
The comparison group and incentivized group had nearly identical average performance on the pretest as shown in Table II (analysis of variance or ANOVA shows p=0.972).  However, the distribution of posttest performance of the incentivized group is significantly better than the comparison group with a p-value of 0.001.  Table II also shows that generally the students with lower performance on the pretest benefit more from the explicit incentive to correct their mistakes in the pretest.

Figure 1 shows the average gain (defined as the arithmetic difference between posttest and pretest scores; gain can therefore range from -100\% to +100\%) vs pretest score (which can range from zero to 100\%) for each student on the four questions repeated from the pretest to posttest in the comparison group and incentivized group. 

 The regions of possible gain are shaded according to posttest score performance categories: green for High posttest performance, yellow for Medium posttest performance and orange for Low posttest performance.  The performance categories are defined as follows: “High,” for scores from 85\% to 100\%; “Medium” for scores from 50\% to 85\%; and “Low” for scores from zero to 50\%.  The researchers agreed upon a 50\% cutoff somewhat arbitrarily for separating Low and Medium categories and 85\% cutoff was chosen so that roughly one third of the students scored in the High performance category on the pretest.

Figure 1 and Table II show that students with poor performance on the midterm exam were likely to benefit from self-diagnosis activities in which they submitted the corrected midterm  solutions for 50\% of the points lost on each problem.  Therefore, the gap between the High and Low performers on the midterm exam was reduced for this incentivized group on the repeated problems on the final exam. On the other hand, for the comparison group, the gap remained, i.e., scores did not substantially improve, and remained diverse.

The data were analyzed by breaking the students into three groups based on their pretest performance as shown in both Figure 1 and Table II. The initially high-performing students from both the comparison and incentivized groups (scoring 85\% and higher on the pretest) generally performed very well on the posttest regardless of the intervention (see Figure 1).  Most of these students who start in the High pretest category stay in that category.  Students who initially performed at a Medium level on the pretest (scoring between 50\% and 85\%) in the incentivized group perform better on the posttest than the corresponding students who were in the comparison group.  In the comparison group, students in the Medium performance category on the pretest were as likely to improve on the posttest (above the horizontal axis in Figure 1A) as they were to deteriorate (below the horizontal axis).  In contrast, in the incentivized group, almost all of the students in the Medium category on the pretest improved on the posttest (see Figure 1B).  

Furthermore, about half of these students in the incentivized group improved as much as possible on the posttest, saturating the boundary for maximal improvement (see Figure 1B). Among the initially Low performing students (pretest scores less than 50), many students in both comparison and incentivized groups improved on the posttest.  However, the degree to which these struggling students performed on the posttest is highly dependent on whether or not they received a grade incentive to improve.  The students in the Low category on the pretest in the incentivized and comparison groups had an average gain of 45.0\% and 16.2\%, respectively (see Table II). In summary, the gains are much larger for the incentivized group, bringing the average of the Low category to the level of the Medium category, and the Medium category to the High category (see Figure 1 and Table II).

\begin{figure}
	\centering
	\includegraphics[width=\linewidth]{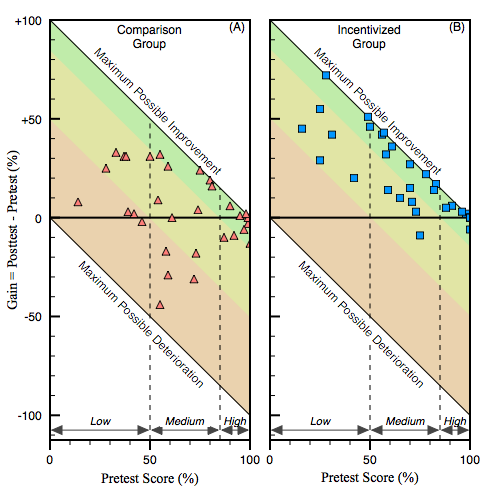}
	\caption{Average gain (defined as the difference between posttest and pretest score) vs pretest score for each student on the four questions repeated from the pretest to posttest in the comparison group (A) and incentivized group (B).  A few students' data points overlap. Red-filled triangles are for each student from the comparison group and the blue-filled squares are for each student in the incentivized group in which students received an explicit grade incentive to correct their own mistakes on the pretest before answering the same questions on the posttest.  Students whose scores improved are above the horizontal axis; students whose performance deteriorated are below the horizontal axis.}
	\label{fig:figure1}
\end{figure}

\section{Discussion}
An easy-to-implement intervention substantially reduced the performance gap between low- and high- achieving students in quantum mechanics.  Explicitly providing grade incentives to correct mistakes positively affected the performance of students with a diverse spectrum of performance in the midterm exam on subsequent solving of the same problems in the final exam.  Our research suggests that even students in advanced physics courses such as quantum mechanics are more motivated to engage with instructional material in a meaningful way if they are provided an explicit incentive~\cite{motivation2,motivation3,motivation1}. Considering the relative ease with which instructors in physics courses at all levels can implement the intervention in which students are given grade incentives to correct and learn from their mistakes, instructors at all levels should consider giving students this opportunity to learn. Asking students to correct their mistakes in several courses may also help students understand the importance of learning from mistakes and the role of appropriate struggle in learning. 

Providing students with incentives to learn from their mistakes may positively impact learning
without a significant additional effort on the part of the instructors and without requiring a substantial initial or recurring investment of resources. 
We find that in quantum mechanics, the corrections themselves take very little time to grade because students seize the opportunity and produce expert-like solutions even when they are given only a few days to correct their own mistakes. We note that while students submitted both their initial and corrected solutions,  explanations of the mistakes they had made and how they corrected them was limited since the instructor (who graded them) did not explicitly tell them that they would be graded on such explanations.

While these results are encouraging, caution is urged in interpreting improvement.  Our findings support the claim that students improve on problems administered a second time when expert-like behavior (self-diagnosing and correcting their mistakes) is explicitly incentivized.  It does not necessarily follow that students have become adept at self-monitoring skills from just two such interventions in the two midterm exams in quantum mechanics. In particular, we compared the performance of students in the incentivized and comparison groups on four other problems (on other topics) on the same final exam for which incentivized group students did not diagnose their mistakes. We find that while the incentivized group scored somewhat higher than the comparison group, the results are not statistically significant on student performance on those four other problems on the final exam.

\section{Summary and Conclusion}
We find that the performance of students in the group in which no incentives were provided shows that while some advanced students performed equally well or improved compared to their performance on the midterm exam on the questions administered a second time, a comparable number of students obtained lower scores on the final exam than on the midterm exam. The wide distribution of students' performance on problems administered a second time suggests that many advanced students do not automatically exploit their mistakes as an opportunity for learning, and for repairing, extending, and organizing their knowledge structure. On the other hand, the performance on the final exam on the same problems suggests that students who were given incentives to correct their mistakes significantly outperformed those who were not given an incentive. The incentive to correct the mistakes on the midterm exam had the greatest impact on the final exam performance on those problems for students who performed poorly on the midterm exam. 

An explicit incentive to correct their mistakes can be an effective formative assessment tool \cite{blackwil}. Offering grade incentives to diagnose and correct mistakes can go a long way to close the performance gap between struggling and high-performing students.  If this type of easy-to-implement intervention is implemented routinely in all physics courses, students are likely to use their mistakes as a learning opportunity and may even develop better self-monitoring skills over time. Considering the relative ease with which instructors in physics courses at all levels can implement the intervention in which students are given grade incentives to correct and learn from their mistakes, instructors should consider giving students at all levels such an opportunity to learn.


\begin{acknowledgments}
We thank the National Science Foundation for award PHY-1202909 and F. Reif and R. P. Devaty for useful discussions.
\end{acknowledgments}

\appendix

\section{An Example Solution}
Below, we provide a student response on problem 1 from the incentivized group to show how a typical student improved from the pretest to posttest.

{\noindent
1) The eigenvalue equation for an operator \op{Q} is given by  $\op{Q}\ket{\psi_i} = \lambda_i \ket{\psi_i} $, with $i = 1...N $.  Find an expression for $\braket{\psi | \op{Q} | \psi}$, where $\ket{\psi}$ is a general state, in terms of $\braket{\psi_i | \psi}$.
}
\vspace*{0.05in}
{\noindent
  \begin{align*}\tag*{Pretest}
   \braket{\psi| \op{Q} | \psi} &=\\
    &= \sum\limits_{i=1}^N \braket{\psi| \lambda_i | \psi_i} \\
    &= \sum\limits_{i=1}^N \braket{\psi| (\braket{\psi_i | \psi}) | \psi_i} \\
    &= \sum\limits_{i=1}^N |\braket{\psi_i| \psi}|^2 
  \end{align*}
  \begin{align*}\tag*{Posttest}
   \braket{\psi| \op{Q} | \psi} &=\sum\limits_{i=1}^N\braket{\psi | \op{Q} | \psi_i}\braket{\psi_i|\psi}\\
   &=\sum\limits_{i=1}^N\braket{\psi | \lambda_i | \psi_i}\braket{\psi_i|\psi}\\
   &=\sum\limits_{i=1}^N \lambda_i \braket{\psi | \psi_i}\braket{\psi_i|\psi}\\
   &=\sum\limits_{i=1}^N \lambda_i |\braket{\psi_i|\psi}|^2
  \end{align*}
}

\end{document}